\documentclass[namedreferences]{solarphysics}

\usepackage[hyperref,optionalrh]{spr-sola-addons} 
\usepackage{graphicx}        
\usepackage{color}           
\usepackage{breakurl}        

\usepackage[margin=2.5cm]{geometry}

\chardef\us=`\_

\begin{document}


\begin{opening}

\title{Arecibo Multi-frequency IPS Observations: Solar Wind Density Turbulence Scale Sizes and their Anisotropy}

\author[addressref={aff1,aff2,aff3},corref,email={mano.rac@gmail.com}]{\inits{P.~K.}\fnm{P.~K.}~\lnm{Manoharan}\orcid{0000-0003-4274-211X}}
\author[addressref={aff4},email={csalter.wfc@gmail.com}]{\inits{C.~J.}\fnm{C.~J.}~\lnm{Salter}}

\address[id=aff1]{Florida Space Institute, University of Central Florida, Orlando, FL 32826, USA.}
\address[id=aff2]{Heliophysics Science Division, NASA Goddard Space Flight Center, Greenbelt, MD 20771, USA}
\address[id=aff3]{The Catholic University of America, Washington, DC 20664, USA}
\address[id=aff4]{Arecibo Observatory (Retired), Puerto Rico 00612, USA.}

\runningauthor{Manoharan et al.}
\runningtitle{\textit{Multi-frequency Interplanetary Scintillation Observations}}

\begin{abstract}

We present an analysis of interplanetary scintillation (IPS) observations conducted 
with the Arecibo 305-m radio telescope during the minimum phase at the end of solar 
cycle 24 and the onset of solar cycle 25. These observations span a broad 
frequency range of $\sim$300 to 3100 MHz, encompassing the P-, L-, and 
S-bands, and covered heliocentric distances from $\sim$5 to 200 solar radii.
Each L-band observation provided simultaneous measurements across a bandwidth 
of approximately 600 MHz. Furthermore, whenever feasible, the near-simultaneous 
measurements of a source acquired across all three frequency bands were useful to 
study the scintillation characteristics over a much wider frequency band along the 
same line of sight through the heliosphere.
The dynamic spectrum of the scintillations obtained at L-band 
shows a systematic decrease in the scintillation index 
from the lowest to the highest frequency, offering valuable insight into the influence 
of the solar wind density microstructures responsible for scintillation.
Analyses of the scintillation index ($m$) for multiple sources at L-band,
along with near-simultaneous observations of selected sources covering the P-, L-, and 
S-bands, clearly demonstrate a wavelength dependence of $m \propto \lambda^\omega$,
which inherently leads to a dependence of $m$ on the Fresnel scale, 
when considering the effective distance to the scattering screen, $z$.
The index $\omega$ ranges between $\sim$1 and 1.8.
The average $\omega$ value of a source, determined from observations made on
different days (i.e., at a range of solar offsets to mitigate the influence of 
possible day-to-day variations in solar-wind turbulence) exhibits variability across 
sources. 
The results on the radial dependence of scintillation agree with earlier IPS measurements. 
The temporal power spectra obtained 
over the wide frequency range exhibit a power-level evolution in accordance with the 
wavelength dependence, and a broadening with increasing observation frequency. 
Furthermore, the increased temporal-frequency rounding of the `Fresnel knee' in the 
spectrum with the observing frequency suggests a novel phenomenon: an increase in 
anisotropy as the scale size of the density-turbulence structure decreases.

\end{abstract}


\end{opening}

\section{Introduction} 

The remote-sensing technique of `interplanetary scintillation' (IPS) is a powerful tool 
for probing the solar wind over a wide range of heliocentric distances, from near the 
Sun out to the Earth's orbit, and both within and away from the ecliptic plane -- regions 
often inaccessible to space missions
(e.g., \citealt{hewish1964}; \citealt{coles1978}; \citealt{kojima1987}; 
\citealt{mano1990}; \citealt{mano1993}; \citealt{shishov2010}).
IPS also offers a simple method to identify compact sub-arcsecond structures in radio 
sources
(e.g., \citealt{little1966}; \citealt{salpeter1967}; \citealt{rao1974}; \citealt{readhead1978}; 
\citealt{mano2009}; \citealt{morgan2019}). 
Routine IPS observations at 327 MHz have been conducted over three solar cycles using 
the multi-antenna system at the Institute for Space-Earth Environmental Research, Nagoya 
University 
(\citealt{kojima1990}; \citealt{tokumaru2012}), 
and the large-steerable Ooty Radio Telescope at the Radio Astronomy Centre of the 
National Centre for Radio Astrophysics, Tata Institute of Fundamental Research 
(\citealt{swarup1971}; \citealt{mano1993}; \citealt{mano2012}). 
The Mexican Array Radio Telescope (MEXART) operates as a dedicated transit IPS telescope 
at a central frequency of 140 MHz (\citealt{americo2022}), while the Big Scanning Array 
of the Lebedev Physical Institute regularly monitors IPS at 111 MHz 
(\citealt{igor2023}).
These systems, operating within the frequency range of approximately $\sim$110 -- 327 
MHz, constitute the Worldwide IPS Stations (WIPSS) network 
(\citealt{bisi2021}),  
which aims to provide standardized IPS data for tomographic reconstruction of the solar 
wind 
(e.g., \citealt{mano2010}; \citealt{jackson2020})
to support and improve space-weather science and forecasting capabilities.

Recently, astronomical facilities operating between 50 and 250 MHz, such as the Low 
Frequency Array (LOFAR) 
(\citealt{fallow2013}) 
and the Murchison Widefield Array (MWA) at the Murchison Radio Observatory 
(\citealt{kaplan2015}), 
have also been employed for IPS-science-based studies. Additionally, IPS observations 
with the MWA telescope have been extensively used to identify and survey compact 
components in radio sources 
(\citealt{rajan2018}).

Numerous IPS studies conducted with various radio telescopes at different 
frequencies have provided significant insights into the large-scale structure and long-term
variations of the solar wind, offering valuable contributions to solar physics, 
particularly in understanding inner heliospheric processes and space weather phenomena 
such as coronal mass ejections (CMEs), solar wind interaction regions, and intense 
interplanetary shock waves that can cause severe geomagnetic storms and disrupt critical 
ground- and space-based technologies 
(e.g., \citealt{bourgois1985}; \citealt{gapper1982}; \citealt{kojima1987}; \citealt{asai1998}; 
\citealt{yama1998JGR}; \citealt{mano2006}; \citealt{andy2006}; \citealt{bisi2010}; 
\citealt{fallow2013}; \citealt{kaplan2015}; \citealt{baron2024}; \citealt{igor2023}). 
Some IPS results have also been validated with {\it in-situ} solar wind measurements 
(e.g., \citealt{colesetal1978}; \citealt{hayashi2003}; \citealt{bisi2009}; \citealt{mano2012}).
IPS findings have contributed to the development of models predicting the propagation of 
space weather phenomena
(e.g., \citealt{mano1995SoPh}; \citealt{mano2006}; \citealt{vrsnak2013}; \citealt{iwai2023}).

Moreover, IPS observations provide insight into the spatial spectrum of solar wind 
electron density fluctuations at scales comparable to the diffraction scale (i.e., the 
first Fresnel zone radius). 
To comprehensively understand the physical processes underlying density turbulence over
different Fresnel scales, multi-frequency measurements, especially simultaneous 
observations, are essential.  This paper presents IPS observations of many 
radio sources from the Arecibo 305-meter radio telescope in three frequency bands: P band 
(302 -- 352 MHz), L-band Wide (1125 -- 1735 MHz), and S-band Wide (2700 -- 3100 MHz).
These bands cover diffraction scales ranging from $\sim$100 to 400 km as functions of 
heliocentric distance. 
The paper is structured as follows.  Section 2 briefly describes the Arecibo system 
used for IPS and data analysis procedures.  Section 3 presents a theoretical overview 
relevant to multi-frequency IPS observations. Sections 4 and 5 present the evolution 
of scintillation within the inner heliosphere during the initial minimum phase of solar 
cycle 25. Section 6 analyzes the temporal-frequency spectrum and addresses the anisotropy 
of solar wind density structures. Section 7 summarizes the key findings.

\section{Arecibo IPS Observations and Data Analysis}

\begin{figure}[h]

\centerline{\includegraphics[width=8.7cm]{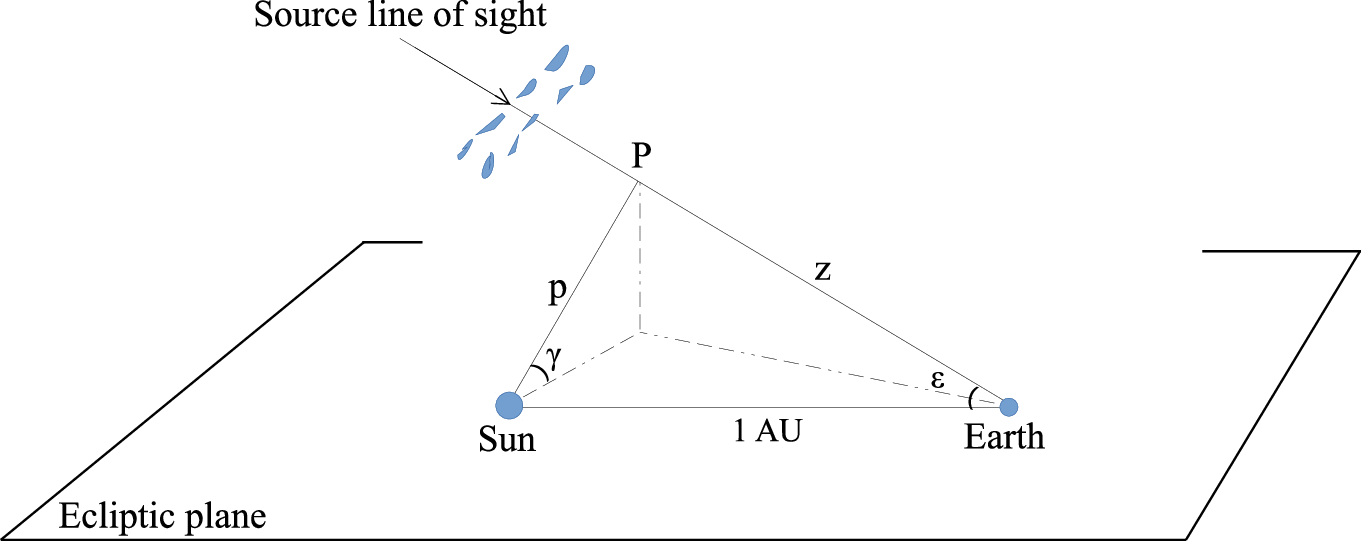} }
\caption{
Geometry of IPS observations. 
As the heliospheric equator nearly coincides with the ecliptic plane, the angle, $\gamma$, 
approximates to the heliographic latitude of the `P' point.} 
\label{fig_1}
\end{figure}

IPS observations reported in this study were taken with the Arecibo 305-m Radio 
Telescope during the end of solar cycle 24 (August--October 2019) and the start of 
solar cycle 25 (March--August 2020), encompassing the minimum phases of both cycles. 
Due to limited observation time in 2019, during the end phase of solar cycle 24, 
the number of observed sources was limited.
However, due to COVID-19 restrictions, the telescope time was available in 2020, and 
frequent observations were possible, with more than ten sources observed daily.  The 
sources were selected from the Ooty IPS list, with scintillating flux densities of 
$\Delta S \ge 1$ Jy at 327 MHz (\citealt{mano2009}; \citealt{mano2012}). 
These observations covered solar elongations ($\varepsilon$) in the range of 
$\sim$1$^\circ$ -- 70$^\circ$, corresponding to heliocentric distances of 
$\sim$5 -- 200 $R_\odot$ (1 solar radius, $R_\odot$ = 6.96$\times 10^5$ km and 1~AU 
$\approx$ 215 $R_\odot$). Several observations out of the ecliptic probed the high 
heliographic latitude solar wind (see Figure \ref{fig_1}).

The Arecibo radio telescope operated over a wide frequency range, from 300 MHz to 10 GHz, 
(a wavelength range of $\lambda \sim 1$~m to 3 cm). This study presents
IPS observations using three receivers: P band (302 -- 352 MHz), L-band Wide
(1125 -- 1735 MHz), and S-band Wide (2700 -- 3100 MHz). All three systems had similar 
high gains of $\sim$10~${\rm K\, Jy^{-1}}$ (\citealt{ao2013}; \citealt{mano2022PSJ}).
The Arecibo telescope covered a declination range of $-1^\circ$ to $+38^\circ$, 
allowing observations of the inner heliosphere, i.e., near-Sun region, out to 1 AU, for 
over six months per year, including the entire summer solstice in the northern 
hemisphere and part of the winter solstice.  The telescope provided a maximum tracking 
time of 2.75 hours at a declination of about +20$^\circ$. 
The tracking time decreased on either side of this declination (\citealt{ao2013}). 

Each source was observed for three minutes, centered within the tracking span for the source's 
declination, thus avoiding the edges of the tracking limits. Immediately afterward, the 
telescope pointing was shifted east of the source in right ascension to observe an 
off-source region for an additional three minutes. This approach ensured that an identical part 
of the dish surface was tracked during both the on-source and off-source scans, helping 
to account for any systematic gain variations in the system. The source deflection 
was determined by calculating the difference between the mean levels of the on-source 
and off-source observations.

Each day, several sources were observed, and multiple days of observations of these
sources sampled various regions of the inner heliosphere. Approximately 1230 
on-source scans were observed using the P-, L-, and S-band systems. However,
nearly 75\% of the scans  were taken with the L-band  system, covering a range of 
heliocentric distances from $\sim$5 to 150 $R_\odot$.  About 20\% of the scans were 
taken with 
the P-band system, focusing on heliocentric distances farther from the Sun, i.e., 
$\ge$40 $R_\odot$. The remaining 5\% of scans were taken with the S-band system, 
covering distances within $\sim$75~$R_\odot$ of the Sun. The lower percentage
of P- and S-band observations was primarily due to the planned prolonged maintenance
in the earlier part of the observing program.

Each L-band observation, covering a broad bandwidth of $\sim$600 MHz, has been valuable 
for studying scintillation variations with observing frequency. Additionally, selected 
sources were observed quasi-simultaneously, covering the three bands within a 
30-minute window and providing insights into scintillation dependence over this wide 
frequency range, at varying distances from the Sun, and for sources with different 
angular sizes. The high sensitivity of the Arecibo telescope enabled the detection of 
even weak scintillating flux densities.

The data were recorded using the single-pixel mode of the FPGA-based Mock spectrometer 
system (see \url{https://naic.nrao.edu/arecibo/phil/hardware/pdev/pdev.html}; 
\citealt{mano2022PSJ}).
The Mock spectrometer consists of seven boxes, each handling an 80-MHz bandwidth. 
For the P-band observations, a single Mock spectrometer box was used to record a  
53.3-MHz bandwidth of dual-polarization data centered at 327 MHz, with each polarization 
divided into 1024 channels and an integration time of 2~ms. For the L and S bands, 
seven and four Mock boxes were utilized, respectively, with each box processing an 
80-MHz bandwidth of dual-polarization data. Each 80-MHz bandwidth was further split 
into 2048 channels per polarization. As at P band, the S-band data were sampled at 2~ms.
However, the L band sampling was increased to 1~ms to prevent aliasing due to radio 
frequency interference (RFI), likely originating from local radar systems.


The bandpass correction was applied to every 10-s block of data
(see \url{https://naic.nrao.edu/arecibo/phil/masdoc.html}). 
Data from each L- and S-band Mock box were further split into 8$\times$10-MHz subbands, with 256 
channels in each.  Individual bad channels within a subband were identified by high 
transient rms values ($>$3$\sigma$) and flagged. For each subband, the frequency-averaged 
spectral densities over the remaining good channels were then obtained, 
resulting in 180-s total-power time series of 56$\times$10 MHz and 
32$\times$10 MHz subbands of data for the L and S bands, respectively. For the P-band 
observations, total-power time series for 5$\times$10 MHz subbands, 
each 
including 256 channels, were similarly computed. The same procedures were repeated for the 
off-source scan data at all frequencies.

In the observed total-power time series of each 10-MHz subband, any slow variations at 
frequencies lower than 0.1 Hz were removed by subtracting running averages of 10-s 
duration.  This running-average subtraction is useful to remove any gradual changes in 
the system gain and the response of the gradually drifting `screen' close to 
the observing plane (e.g., the ionosphere layer). The temporal power spectrum for 
each subband was calculated by Fourier transforming the 30-s data blocks, separately for 
each polarization. The averaged temporal spectrum of two polarizations was then used for 
further analysis (e.g., \citealt{mano2022PSJ}).

\section{IPS Observations and Solar Wind Studies}

The IPS phenomenon arises when
the radiation from a distant compact source (e.g., a radio galaxy or quasar) passes through 
the solar wind's density irregularities ($\delta n_e$). For a sufficiently small source 
size (angular size, $\Theta \lesssim$~400~mas), these irregularities are illuminated coherently, 
causing scattered radio waves to interfere and create a random diffraction pattern on the 
ground. The radial flow of the solar wind translates this pattern into temporal fluctuations 
of the source intensity, known as IPS 
(e.g., \citealt{hewish1964}; \citealt{coles1978}; \citealt{mano1993}).

Extensive literature exists on the theory and applications of IPS observations for solar 
wind studies. Some key references include 
\citet{salpeter1967}, \citet{young1971}, \citet{marians1975}, 
\citet{rumsey1975},  \citet{coles1978}, \citet{readhead1978}, \citet{shishov1978}, 
\citet{uscinski1982}, \citet{hewish1989}, and \citet{mano1993}.
This paper provides a brief overview of the IPS theory necessary to understand the 
observed multi-frequency features of solar wind electron density fluctuations.

\subsection{Background Theory -- Geometry of IPS Observation}

Figure 1 illustrates a simplified geometry of IPS observations. 
The solar elongation, $\varepsilon$, is the angle between the 
Sun-Earth line and the line of sight to the radio source. This angle changes by 
approximately one degree per day due to the orbital motion of the Earth as the radio 
source appears to approach or recede from the Sun. 
IPS measurements represent the integrated effects of the solar wind along 
the total line of sight. However, the solar wind contribution is dominated by the region
around the closest approach point of the line of sight, referred to as the `P' point, 
at a distance ${\rm p } = \sin({\varepsilon})$ in AU (see Figure \ref{fig_1}).  This 
dominance occurs because of the concentrated density turbulence, $\delta n_{e}^{2}(R)$, 
around the region of the `P' point, as on either side of this along the line 
of sight, it decreases even steeper than 
$\delta n_{e}^{2}(R) \propto R^{-4}$ (e.g.,  \citealt{armstrong1978}; \citealt{mano1993};
\citealt{asai1998}; also see subsequent sections).  
The results on the radial dependence of the scintillation 
power obtained from this study during the initial minimum phase of the current solar 
cycle are discussed in Section 5.


\subsection{Scintillation Index -- Radial Dependence}

The IPS of a source is quantified by its scintillation index, \(m\), which is the rms of the 
intensity fluctuations normalized by the average intensity, 
$m\, =\, \{ \langle \delta I(t)^2 \rangle \}^\frac{1}{2}/\langle I \rangle $
(e.g., \citealt{hewish1964}; \citealt{mano1993}). Since the rms of intensity fluctuations 
is equivalent to the square root of the integrated temporal power spectrum, the scintillation 
index \(m\) can also be estimated by,
$m = \left( \int P(f) \, {\rm d}f \right)^{\frac{1}{2}}/{\langle I \rangle}$.
\(P(f)\) is derived from the Fourier transform of the time series of intensity fluctuations, 
\(I(t)\). 
The power level of the temporal spectrum 
is determined by the scattering strength, \(C^{2}_{\delta n_e}(R) \propto R^{-\beta}\), which 
is integrated along the line of sight. With the scattering radial index, \(\beta\), typically 
ranging from 4 to 4.5 (e.g., \citealt{armstrong1978}; \citealt{mano1993}; 
\citealt{asai1998}; \citealt{mano2012}), 
the dominant contribution to scintillation arises from the solar wind 
plasma layers near the point of closest solar approach along the line of sight (see Figure 
\ref{fig_1}).

The rms of intensity fluctuations, $\delta I(t)(R)$, increases as the Sun is approached. 
For a perfect coherent point source, the index, {\it m}, reaches a maximum value close 
to unity at a heliocentric distance, $R_{peak}$, and remains nearly at the same level
(i.e., the saturated level) for closer distances. The distance of the saturation 
onset, $R_{peak}$, depends on the frequency of observation, and it moves closer to the Sun 
with increasing frequency of observation (\citealt{marians1975}; also see Section 5.2). 
For a finite source size, the index maximizes at $R_{peak}$, although $m_{peak}$ is always 
less than unity for a source of finite angular size.  The $m$ 
then decreases for smaller solar offsets, due to the incoherency of scintillation caused 
by the source structure.  As the angular scale size of the source increases, $m_{peak}$ 
decreases and completely vanishes, when the source's angular size is much greater than the 
Fresnel scale. The maximum value of the index, $m_{peak}$, or the radial-dependence of the 
$m(R)$ curve, provides an estimate of the source size 
(e.g., \citealt{marians1975}; \citealt{mano1993}). 

At distances $R~<~R_{peak}$, the scattering is `strong'.
In contrast, at $R > R_{peak}$, the scattering is `weak', 
and the index $m^2$ is linearly related to $\delta  n_{e}^{2}$. The `{\it m -- R}' curve
provides the radial dependence of the density fluctuations, $\delta  n_{e}(R)$ (see
Section 5).

\subsection{Model IPS Temporal Spectrum}


Under weak-scattering conditions, for a thin layer of solar wind plasma located at 
the closest solar approach plane 
perpendicular to the line of sight at a distance {\it z} from the observer, 
(i.e. on the {\it x-y} plane at the `P' point),
the 
contribution to the temporal spectrum includes an integral over the {\it x-y} plane,
 expressed as:
\begin{eqnarray}
P(f,z)_{R = p} = \frac{(2\pi r_e \lambda)^2}{|V(z)|}
       \int^{+\infty}_{-\infty} {\rm d}\kappa \;\; 
       C^{2}_{\delta n_e}(R) \cdot \Phi_{\delta {n_e}}(\kappa_x, \kappa_y,z)
       \cdot F_{diff} (\kappa,z) \nonumber \\ \;\; \cdot F_{source} (\kappa,z,\theta_o) ,
\end{eqnarray}
where \(\lambda\) is the observing wavelength, \(r_e\) is the classical electron 
radius, and $\kappa$ is the wavenumber (\citealt{armstrong1990}; \citealt{mano1990}).
The scintillation is controlled by the angular size of the source, 
$F_{source} (\kappa,z,\theta_o)$, the diffraction term, $F_{diff} (\kappa,z)$ 
(Fresnel-filter term), scaled by the radial velocity of the solar wind, $V(z)$, 
perpendicular to the signal path in the $z$ direction and the observation 
wavelength $\lambda$ (frequency,  $\nu$).
The resultant temporal spectrum, $P(f)$, is the sum of contributions of plasma layers 
between the observer and the source along the {\it z} direction, as given by:
\begin{eqnarray}
P(f) = \int_{observer}^{\infty} P(f,z){\rm d}z.
\end{eqnarray}

The shape of the temporal spectrum reflects the spatial spectrum of intensity turbulence, 
\(\Phi_{\delta {n_e}} (\kappa) \propto \kappa^{-\alpha}\). 
The diffraction term, $F_{diff} = \sin^2\left({\kappa^2}/{\kappa^{2}_{diff}} \right) =
\sin^2\left({\kappa^2\lambda z}/{4\pi}\right) $, 
influenced by the observing wavelength, defines the Fresnel scale involved in scintillation. 
It introduces a knee-like feature around the Fresnel wavenumber, 
\(\kappa_{diff} = \sqrt{4\pi/{\lambda z}}\). The corresponding temporal frequency, 
\(f_{diff} =  V(z)\,\kappa_{diff}/2\pi\), shifts to higher values as the solar wind radial 
velocity increases.  For an isotropic distribution of density irregularities, the wavenumber 
\(\kappa\) is defined as 
\(\kappa = (\kappa_x^2 + \kappa_y^2)^{\frac{1}{2}}\), where {\it x} and {\it y}
represent the radial and perpendicular directions, respectively. 

In general, the scales of solar wind density turbulence are influenced by the radial 
expansion of the solar wind and the orientation of the mean magnetic field. If present, 
such anisotropy modifies the spectrum to
\(\Phi_{\delta n_e}(\kappa) \propto \left (\kappa_{x}^{2} + \frac{\kappa_{y}^{2}}{{\rm AR}^2} \right )^{-\alpha /2}\), 
where AR is the axial ratio of the density irregularities 
(e.g., \citealt{armstrong1990}; \citealt{colesharmon1991}; \citealt{grall1997}; 
\citealt{yama1998JGR}). 
An AR value greater than one results in the rounding of the Fresnel knee of the spectrum 
(e.g., \citealt{scott1983}; \citealt{mano1987}; \citealt{mano1994}). 
The effect of the dissipative scale, $S_i$, is observed as a sharp decline in power at the 
high-frequency end of the temporal spectrum, \(\exp(-\kappa^2/\kappa_{i}^{2})\), 
where $\kappa_i = 3/S_i$
represents the inner-scale cutoff 
(e.g., \citealt{mano1987}; \citealt{colesharmon1991}; \citealt{yama1998JGR}); 
\citealt{mano2000}; \citealt{spangler2002}).

Both the observing bandwidth and the finite angular size of the source result in a decrease 
in source coherence, leading to a reduction in the overall level of scintillation, $m$~$<$~1
(e.g., \citealt{little1966}; \citealt{mano1993}). 
The Fourier transformation of the brightness distribution of a source having a finite angular size, 
i.e., the visibility function of the source, $\exp(-\kappa^2 z^2 \theta_o^2)$ 
attenuates scintillation at high wavenumbers above $\kappa_{source} = (z\theta_o)^{-1}$, 
where $\theta_o$ is the angular radius of the source at the $e^{-\frac{1}{2}}$ level
(the full width at half maximum of the source brightness distribution,
${\Theta}\,=\,2.35{\theta_o}$)
(e.g., \citealt{marians1975}; \citealt{coles1978}; \citealt{mano1994}; \citealt{yama1998JGR}). 
The Gaussian cutoff of the source size,  \(\theta_o \ll \sqrt{\lambda/z}\), occurring at 
frequencies $f_{source}\,=\,{\rm V_{sw}}/2\pi(z\theta_o)$, primarily affects the high-frequency 
portion of the temporal spectrum. 
However, a source size of \(\theta_o \gtrsim \sqrt{\lambda/z}\) can considerably lower the 
Fresnel knee of the spectrum, resulting in reduced power of the scintillations.

By fitting a multi-parameter model based on Equation (2) to the observed temporal 
spectrum, it is possible to derive the typical speed of the solar wind and the power-law 
characteristics of the turbulence spectrum (e.g., \citealt{mano1990}; \citealt{mano1994}).
When the signal-to-noise ratio of the power 
spectrum is significant in the high-frequency region, it can be useful for estimating  the 
combined effect of the angular size of the radio source and the inner-scale size of the
turbulence. Conversely, if the angular size of the source is known from the Very Long 
Baseline Interferometry (VLBI) observations, 
this spectral-fitting approach can provide estimates of the dissipative (or inner) scale 
of the turbulence  (e.g., \citealt{mano1987}; \citealt{mano1994}; \citealt{mano2000};
\citealt{yama1998JGR}).

\section{Multi-frequency IPS -- Results and Discussion}

\subsection{Frequency Dependence of Scintillation  -- Dynamic Spectrum}

\begin{figure}[t]
\centerline{\includegraphics[width=6.5cm,angle=-90.0]{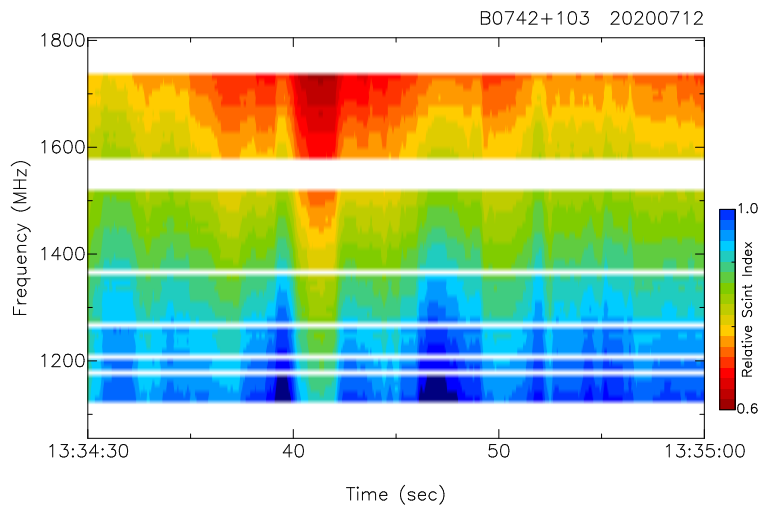} }
\caption{
Scintillation dynamic spectrum of B0742+103 observed with the Arecibo L-band system. 
The large white patch between 1520 and 1580 MHz is the gap in the L-band system and
other smaller gaps at lower frequencies are due to interference.
}
\label{fig_2}
\end{figure}

The wide frequency coverage of the IPS observations in the P-, L-, and S-bands ($\sim$300
-- 3100 MHz) using the Arecibo telescope enables analysis of the frequency dependence of 
scintillation.  Notably, the IPS recordings in the L-band, covering a broad range of 
$\sim$600 MHz, enable the investigation of scintillation variations in 10-MHz intervals.
Figure \ref{fig_2} shows the L-band dynamic spectrum of IPS for the source 
B0742+103, observed 
on 12 July 2020. The solar elongation of the source was 12$^\circ$, corresponding to 
$R$~$\approx$~45 $R_\odot$. This plot presents a 30-s time series, after applying a 10-s 
running mean subtraction (i.e., 40-s data was included in the analysis). For each 
frequency channel, the scintillation index was computed from the rms of a 200-sample 
block (i.e., 200-ms data). 
The source deflection, $\langle I \rangle$, represents the difference between the mean levels of 
the on-source and off-source observations over the 30-s data. 

In Figure \ref{fig_2}, a prominent feature is the systematic decrease in the scintillation 
from the lower to the higher frequencies, 
illustrating the direct proportionality between scintillation and wavelength. 
Additionally, the random temporal variability of scintillation for each frequency band is evident 
along the time axis.
The reduction in scintillation with increasing frequency is consistent with dynamic spectra observed 
in LOFAR data and numerically simulated spectra (\citealt{fallow2013}; \citealt{coles1984}). However,
such a systematic reduction was not clear when the observation covered the weak-to-strong 
scattering transition (e.g., \citealt{coleslee1980}). 

The dynamic spectrum provides valuable insights into the influence of solar wind density 
microstructures that generate scintillation. As outlined in the following section, the frequency 
(or wavelength) dependence of scintillation is clearly revealed when examining the scintillation 
index, \(m(\lambda, R)\), for each 10-MHz channel of the Arecibo bands.

\begin{figure} 
\centerline{
\includegraphics[width=12.4cm]{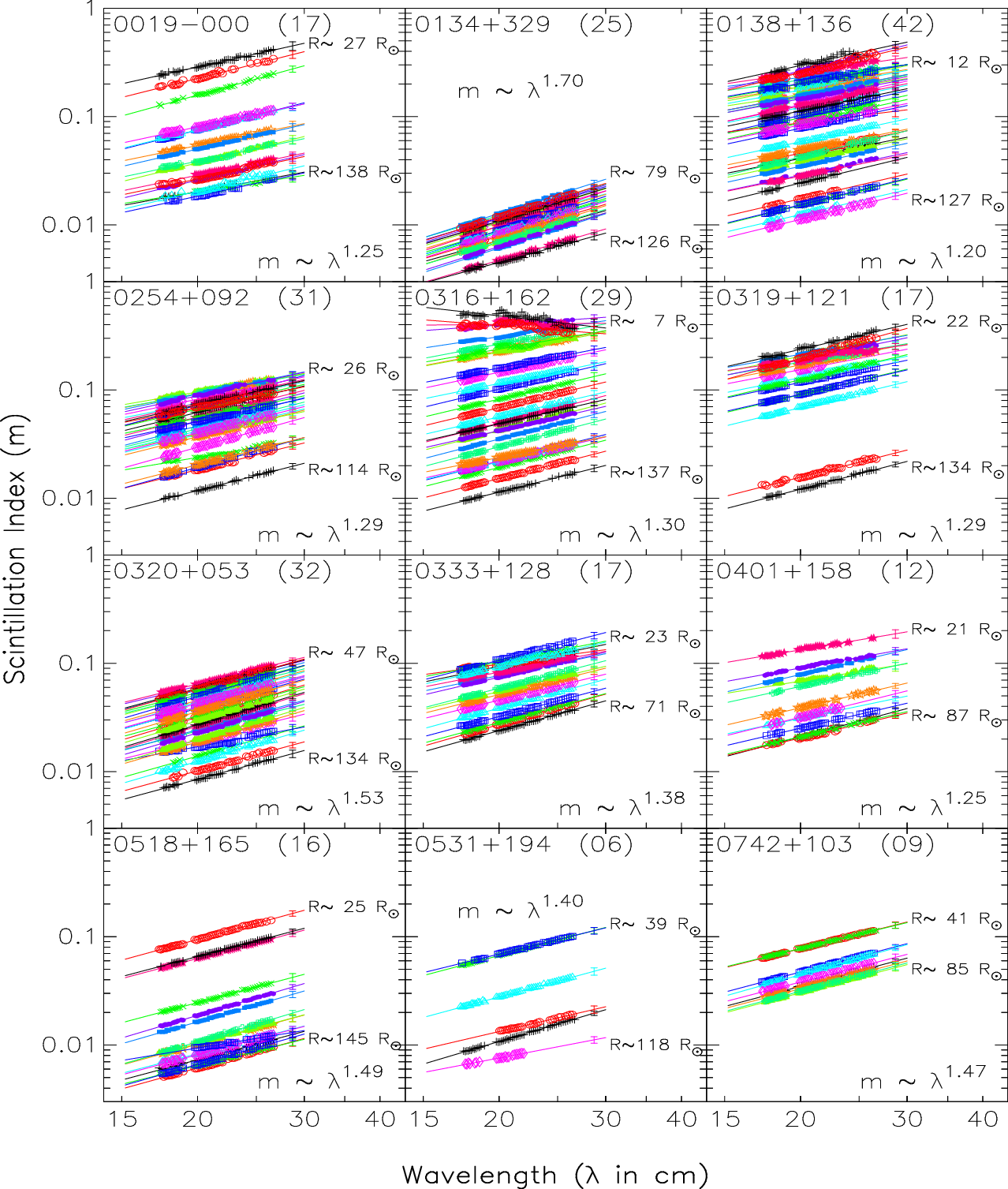} }
\caption{
The `$m$ -- $\lambda$' plots of multiple sources observed at L-band system. 
In each plot, {\it m} values measured on a day across the band are plotted by a distinct color, 
with the corresponding best-fit line.
A vertical bar plotted at $\lambda$ = 29 cm on each day plot is the typical average error 
on the measured indices at the $\pm$1-$\sigma$ level. The number of days observed for each source 
is given in parentheses to the right of the source name.
}
\label{fig_3}
\end{figure}

\subsection{L-band System -- `$m$ -- $\lambda$' Plots}

Figure \ref{fig_3} displays log-log plots of the scintillation index, $m$, as a function 
of wavelength,
$\lambda$, across 56 channels, each with a 10-MHz bandwidth, from the L-band system. 
The figure includes data for 12 sources observed between March and August 2020. 
During this period, which marked the early minimum phase of solar cycle 25, solar 
activity was relatively quiet.
These sources have scintillating flux densities, $\Delta S$~$>$~1 Jy, at 327 MHz 
(\citealt{mano2009}; \citealt{mano2012}).
The number of days of observations for each source ranged from 6 to 42, as indicated 
in parentheses next to the source name. Additionally, the minimum and maximum solar 
offsets for each source are shown, corresponding to the top (high $m$ day) and the bottom 
(low $m$ day) curves, respectively. For instance, the closest solar offset 
observation for source 0316+162 was at 7~$R_\odot$ in the strong scattering region, 
while the maximum solar offset was 137~$R_\odot$.

In the strong scattering regime, the $m - \lambda$ relationship differs from that 
in the weak scattering regime, which occurs at solar offsets $>$15 $R_\odot$. 
Specifically, at L-band at a central wavelength of $\lambda$ = 21\,cm (frequency, 
$\nu$ = 1420~MHz), the scintillation index peaks around 10 -- 15 $R_\odot$ (see 
Figure 8). 
Interestingly, at the solar offset of $\sim$7~$R_\odot$, on the longer wavelength 
side, the scintillation index reaches or crosses the turnover point, $R_{peak}$, 
and shows a decrease (refer to the plots of B0316+162 and B0138+136 in 
Figure \ref{fig_3}). On the shorter wavelength side, the index approaches the 
turnover region, and the reduction in scintillation is significantly less than at 
the longer wavelengths. This wavelength dependence in the strong scattering region 
reveals intriguing characteristics, offering insights into the turbulence spectrum 
in strong (or saturated) scattering conditions and this will be dealt in a separate
study.

\begin{figure} 
\centerline{
\includegraphics[width=4.5cm,angle=-90]{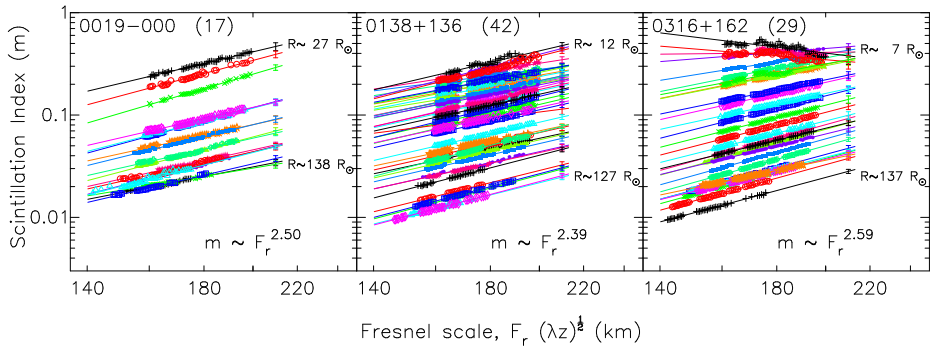} }
\caption{ 
Scintillation index plotted against the Fresnel scale for three very compact 
sources shown in Figure \ref{fig_3}. The format and color scheme are similar 
to Figure \ref{fig_3}.
A vertical bar plotted at F$_{\rm r}$ = 210 km on each day plot is the average error 
on the measured indices at the $\pm$1-$\sigma$ level.  }
\label{fig_4}
\end{figure}

The IPS observations presented in Figure \ref{fig_3} include a range of heliocentric 
distances. It reveals that for a given source, 
the `$m$ -- $\lambda$' curves at different distances in the weak-scattering regime 
exhibit similar slopes. 
For each source, the 
average slope, $\omega$, ($m~\propto~\lambda^\omega$) obtained at heliocentric distances 
$>$20~$R_\odot$, is indicated. The index, $\omega$, ranges between 1.2 and 1.7. 
A higher index, i.e., a steeper `$m$ -- $\lambda$' curve, signifies a sharp drop in 
scintillation at shorter wavelengths or a steeper increase of $m$ at longer wavelengths. 
The day-to-day variation in $\omega$ values observed for a given source at different 
solar offsets can be primarily attributed to changes in $\delta n_{e}$.


Multi-wavelength IPS studies hold significant potential in the study of turbulence 
micro-scale structures.
Since scintillation phenomena are governed by the Fresnel scale, $\sqrt{\lambda z}$,
comparing scintillation as a function of the Fresnel scale can provide valuable insights into the scintillation scales at different observing wavelengths. This approach can also be beneficial for broad inter-comparisons of IPS observations from different radio sources at various frequencies.
However, in line-of-sight integrated IPS observations, the distance to the scattering screen, $z$, is not a fixed parameter. The scattering screen can involve a finite thickness and is effectively determined by the solar wind layers located near the point of closest solar approach, the `P' point (as illustrated in Figure \ref{fig_a1}). At smaller solar offsets, $\varepsilon$ $<$ 45$^\circ$, the maximum level of scattering is observed at $z=\cos(\varepsilon)$.
Considering this caveat, the $m$ values for three sources, namely B0019-000, B0138+136, and B0316+162, are replotted in Figure \ref{fig_4}, incorporating the effective $z$ at the `P' point for each observation. For observations at large elongations, the effective scattering layer moves closer to the observer, with $z$ $<$ 1 AU. Consequently, the Fresnel scale associated with the effective screen distance becomes smaller.
However, the `$m$ -- $\lambda$' and `$m$ -- $\sqrt{\lambda z}$' plots look nearly identical, and the slopes are also basically the same when the $\sqrt{\lambda}$ is considered (refer to Figure \ref{fig_3}).

\begin{figure} 
\centerline{
\includegraphics[width=5.5cm,angle=-90.]{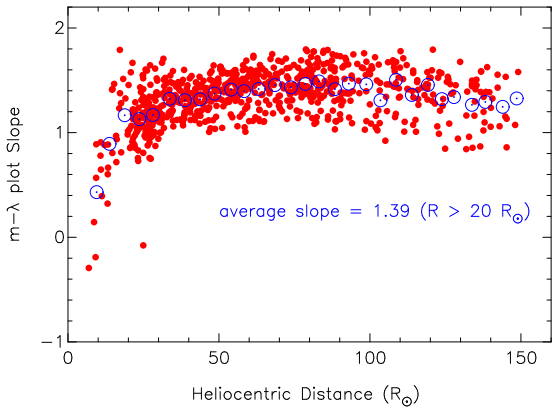} }
\caption{
Arecibo L-band IPS observations: The slope of `$m$ -- $\lambda$' curve plotted as a function 
of heliocentric distance.
}
\label{fig_5}
\end{figure}

\subsubsection{The L-band System -- Heliocentric Distance Dependence of `$m$ -- $\lambda$'}

The IPS observations with the Arecibo L-band system included a large number of sources in 
the heliocentric distance range of $\sim$6~--~150~$R_\odot$. Figure \ref{fig_5} shows 
the slope, $\omega$, of the `$m$ -- $\lambda$' dependency plotted against the heliocentric 
distance. In the strong scattering
regime, the slopes are flatter than the weak-scattering curves. Whereas at distances 
$>$20~$R_\odot$, slope values range between $\sim$1 and 1.8, with an average of 
$\sim$1.39$\pm$0.2, which is consistent throughout the distance range covered.
The 5-$R_\odot$ bin-averaged points are over plotted on the same figure and are shown as 
blue circles.
We observe a mild peak around 80 $R_\odot$. The cause of this effect is currently 
unclear and will be investigated in detail in a future study.


\begin{figure}[t]
\centerline{\includegraphics[width=7.5cm,angle=-90.0]{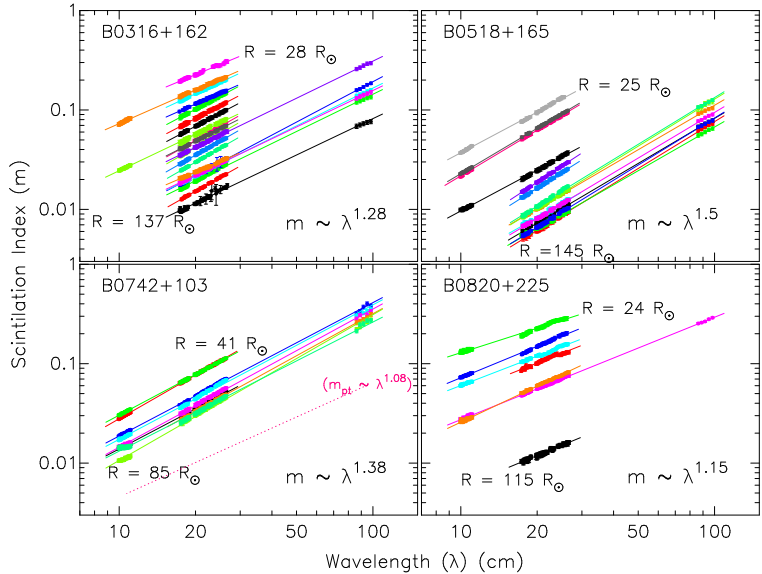} }
\caption{The `$m$ -- $\lambda$' plots of four sources observed with the Arecibo P-, L-, 
and S-band systems.  
As in Figures \ref{fig_3} and \ref{fig_4}, each day's measured {\it m} values are plotted
with a distinct color, along with their corresponding best-fit line.
The dotted line shown along with curves of B0742+103 is the 
`$m_{pt}$ -- $\lambda$' curve of an ideal point source, obtained from the model spectra
using Equation (2), for $\alpha$ = 3.3, $\varepsilon$ = 40$^\circ$, AR =1, and 
$V_{sw}$ = 400 km~s$^{-1}$. 
}
\label{fig_6}
\end{figure}

\subsection{Three Bands -- `$m$ -- $\lambda$' Dependence }

For some of the sources, the IPS have been monitored near simultaneously, i.e., within about
30 minutes, at P-, L-, S-band systems. Figure \ref{fig_6} displays the `$m$ -- $\lambda$' plots of 
four sources, B0316+162, B0518+165, B0742+103, and B0820+225. The format of this figure is 
similar to Figure \ref{fig_3}. 
For the sources B0316+162 (CTA 21, a well-known compact quasar of angular size, 
$\Theta \approx 30~$mas) and B0518+165 (3C138, a compact quasar of size, 
$\Theta \approx 90~$mas), the $\omega$ indices are nearly the same as 
L-band observations (Figure \ref{fig_3}). 
In contrast, for B0742+103, the average index obtained from the 
three-band observations is $\omega_{_{PLS}}$ = 1.38, which is flatter than the average index 
obtained with L-band alone, $\omega_{_{L}}$ = 1.47. 
B0742+103 is a GHz-peaked spectrum radio quasar at a high redshift of 2.624 (\citealt{kharb2010}).
Most of its flux density is contained in a core of size less than 60~mas. Comparison of its 
`$m - R$' dependence at the three-frequency bands suggests that the percentage of the 
scintillating flux at S-band is likely higher than at P-band, and it is consistent with
the flux density peaking around 2.7 GHz. This likely leads to a flatter $\omega$ curve.
BL Lacertae object B0820+225, located at a redshift of 0.951, exhibits a 5 GHz flux density 
of 1.6 Jy. Extensive imaging observations of this source have been made at various frequencies 
using the Very Long-Baseline Array (VLBA) and 
VLBI, specifically in the frequency range of 1.6 -- 15 GHz. Approximately 30--40\% of its flux 
density is associated with an elongated jet-like structure, measuring less than or about 30~mas 
(e.g., \citealt{pushkarev2001}; \citealt{gabuzda2001}).  Over 300 interplanetary 
IPS observations were taken at 327 MHz using the Ooty Radio Telescope, spanning a range of 
solar elongations across different solar cycle phases. These observations revealed an average 
total flux density of approximately 4 Jy and a scintillation flux density of around 1.2 Jy
(\citealt{mano2009}; \citealt{mano2012}). The average $\omega$ value of this source is flatter 
than the above-mentioned sources.

For reference, model spectra at the middle of the P, L, and S bands were computed using 
Equation (2) for both an ideal point source and a source with a size of 60~mas. 
These spectra are displayed in Figure \ref{fig_a2}. For the point source, an $\omega$ 
value of 1.08 was obtained. The corresponding curve is displayed along with the B0742+103 
data in Figure \ref{fig_6}. The difference in power level between the point source spectrum 
and the finite source size spectrum progressively increases with decreasing wavelength (or 
increasing frequency). This results in an $\omega$ value of approximately 1.4 for the source
size of 60~mas, which is consistent with the above results.

Earlier multi-frequency IPS studies have indicated the reduction of scintillation with 
frequency (e.g., \citealt{gapper1981}; \citealt{scott1983}; \citealt{coles1984}; 
\citealt{bourgois1985}; \citealt{andy2006}; \citealt{fallows2006}; 
\citealt{liu2010}; \citealt{fallow2013}; \citealt{morgan2017}).
 
The observed range of $\omega$ values is consistent with the theoretically predicted dependences of $\omega$ = 1.45 and $\omega$ = 1.25, obtained by \citet{armand1987}, respectively, corresponding to density turbulence described by a Kolmogorov spectrum with $\alpha$ = 11/3 and a less steep spectrum with $\alpha$ = 3. As the spectral power decreases more rapidly with increasing temporal frequency, the value of $\omega$ also tends to increase.

Spacecraft observations have shown that the spectral index, $\alpha$, evolves with distance 
from the Sun. In the near-Sun region  ($R$ $<$ 16 $R_\odot$), for the density-irregularity 
scale sizes considered in the present IPS observations, an average $\alpha$ $\approx$ 3 has 
been observed. This steepens to a range of approximately 3.3 -- 3.4 at distances 
between $\sim$20 and 100 $R_\odot$. At larger solar offsets, $R$ $>$~100~$R_\odot$, the 
spectral index $\alpha$ typically reaches a value around 3.7 (e.g., 
\citealt{woo1979}; \citealt{yakovlev1980}; \citealt{armand1987}; \citealt{celniker1987}).
Average $\omega$ values in this study are based on observations at heliocentric distances 
between 20 and 150 $R_\odot$. While the full range was considered, the most of 
observations were concentrated at distances below 100 $R_\odot$. Specifically, at 
frequencies of L band and higher, only compact sources produce measurable scintillation
beyond 100 $R_\odot$. When day-to-day solar wind changes are mitigated, the effect of 
source size on the average $\omega$ cannot be disregarded.


\begin{figure}[t]
\centerline{\includegraphics[width=7.0cm,angle=-90.]{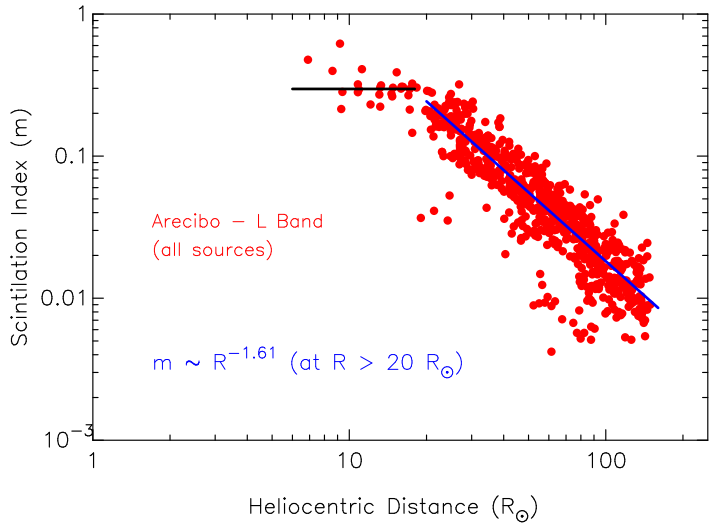} }
\caption{The `$m$ -- $R$' plot of all sources observed with the Arecibo L-band system.
The solid `blue' line represents the straight-line fit to data points at distances 
$>$20~$R_\odot$, while the horizontal black line indicates the average value of $m$, 
approximately 0.3, for distances $\leq$15~$R_\odot$.
}
\label{fig_7}
\end{figure}

\section{Radial Dependence of Scintillation}
\subsection{L-band Observations}

Figure \ref{fig_7} displays the plot of `$m$ -- $R$' of the L-band observations for all 
sources observed, covering the distance range of $\sim$6 -- 150~$R_\odot$.  For each 
source scan in a day, the mean scintillation index has been obtained by averaging over 
the available 10-MHz channels. Thus, it represents the scintillation index at the 
middle of the L-band, $\sim$1420 MHz (refer to Figures \ref{fig_3} to \ref{fig_5}). 
The scintillation index increases as the source approaches the Sun (refer to Section 3.1). 
The saturation level of scintillation, i.e., the onset of the strong scattering region, 
is expected at distances $\lesssim$20 $R_\odot$. Since there are fewer points for 
smaller solar offsets, and these include sources of different angular sizes as well 
as reduced level of $\delta n_{e}^{2}$ in the high-latitude coronal hole regions, the 
turnover of the scintillation index is not seen clearly. For example, in high-latitude 
regions, the 
level of scintillation at a given distance will be less than the level observed at the 
same distance in the equatorial or low-latitude regions (e.g., \citealt{mano1993}; 
\citealt{imamura2014}; \citealt{coles1996}). At distances $\leq$15 $R_\odot$, the 
$m$ values range between 0.2 and 0.6 and the average ($m$ $\approx$ 0.3) is shown by a 
horizontal black line.



The least-square straight-line fit to the $m$ values, at distances $>$20~$R_\odot$, 
provides a relationship, $m~\approx~30\times{\rm R}^{-1.6}$. This is shown as a continuous 
blue straight line in Figure 7.
The `$m$ - R' dependence agrees with the results obtained from the 327-MHz IPS 
observations, made with the Ooty Radio Telescope, in the distance range of $\sim$40 -- 
215~$R_\odot$, for the minimum phases of solar cycles 20 to 24 (\citealt{mano1993}; 
\citealt{mano2012}) and other earlier IPS observations (e.g., \citealt{armstrong1978};
\citealt{asai1998}).

\subsection{Three Bands -- `$m$ -- R' Dependence }

\begin{figure}[t]
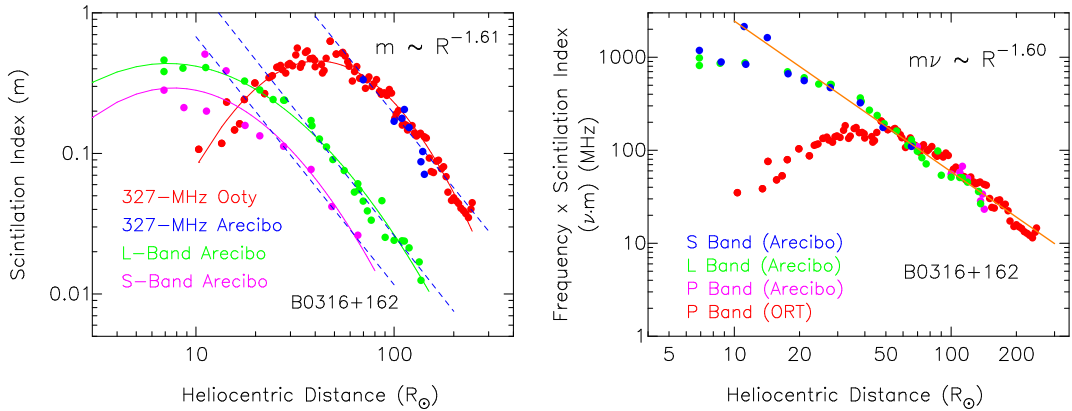

   \centerline{\hspace*{0.00\textwidth}
               \includegraphics[width=6.0cm,angle=-90.0]{figure_8A_new.ps}
               \hspace*{-0.05\textwidth}
               \includegraphics[width=6.0cm,angle=-90.0]{figure_8B_new.ps} 
              }
   \vspace{0.02\textwidth}
\caption{
(Left) The `$m$ -- $R$' plot of the compact source B0316+162, corresponding to the center 
frequencies of Arecibo P- L-, and S-band systems for the year 2020. For comparison, $m$ values
observed at 327-MHz for the year 2018 with the Ooty Radio Telescope (ORT) are also plotted. 
(Right) The scintillation indices, $m$, multiplied by their corresponding 
observing frequencies ($m\cdot\nu$) are plotted. 
In both plots, the least-square fits to the data points in the weak-scattering region are shown as 
straight lines
(also refer to Figure \ref{fig_9} (right), which shows `$m/\lambda^\omega$ -- $R$' plot). }
\label{fig_8}
\end{figure}

Figure \ref{fig_8} (left) shows an example of the `$m$ -- R' plots of B0316+162 (CTA 21) at 
the three Arecibo bands. As mentioned above, the observations at P  and 
S bands were limited in 
number, and only a few sources were covered at all three bands. Moreover, since the 
Arecibo P-band observations did not include distances in the turnover region (i.e., at the 
onset of strong scattering), 
the {\it m} values of B0316+162, measured using the Ooty Radio Telescope at 327 MHz 
during the declining phase of solar cycle 24 in 2018, are plotted for comparison.
The Arecibo points agree with the Ooty observations for the overlapping distance range.

The turnover of scintillation at 327 MHz is observed at $\sim$45~$R_\odot$. Whereas at 
L-band, it shifts to about 15 -- 20~R$_\odot$.
Since the number of observations at S-band is limited, and the turnover is not clear, 
we can only say that
it is at $\lesssim$10~$R_\odot$. 
The least-square fitting to the weak-scattering regions at the three bands provide 
an essentially identical radial dependence of $m \propto R^{-1.6}$.  The analytical solution of 
the line-of-sight integral, $\delta n_{e}^{2}(R)~\propto ~R^{-S}$, leads to a radial
index $S$~=~(2$\times$1.6)+1, resulting in $\delta n_{e}^{2}(R)~\propto ~R^{-4.2}$.

As the source B0316+162 approached the Sun, the `P' points probed the high-latitude 
southern polar region of the heliosphere, at a heliographic latitude of about $-$80$^\circ$~S. 
During the minimum phase 
of the current cycle, the high-latitude regions on the Sun were dominated by a large 
coronal hole of low density 
(see the Carrington map, CR2230, in the 193 \AA\ channel of the Atmospheric Imaging Assembly 
(AIA) on board the Solar Dynamics Observatory (SDO) provided at 
\url{https://sdo.gsfc.nasa.gov/data/synoptic/}).
The low-density turbulent solar wind emanating from 
this open-field region tends to move the onset of the strong-scattering region towards the
Sun. Considering the above distance dependence of $\delta n_{e}^{2}(R) \sim R^{-4.2}$, 
a given level of scattering strength at the high-latitude region is observed at a 
distance closer than that of the equatorial region.
This is consistent with the results of earlier remote-sensing studies
(\citealt{mano1993}; \citealt{coles1996}; \citealt{mano2012}; \citealt{imamura2014}).

The multi-frequency IPS technique employed in this study, especially at higher 
frequencies, allows us to overcome the limitation imposed by Fresnel filtering
at meter-wavelengths.
It enables the investigation of low-frequency temporal variations in 
${\delta n_{e}^{2}}(R,z)$, as close as $\sim$10 $R_\odot$ at S-band, and provides 
valuable insights into the density turbulence.

\subsection{Radial Dependence of Frequency-Scaled Scintillation Index}

Figures \ref{fig_3} to \ref{fig_6} demonstrate the invariance of the `$m$ -- $\lambda$' 
dependence with the distance from the Sun. If the angular size of the source does not 
vary with the observing frequency, $\nu$, the product $m\cdot\nu$ is expected to yield 
a frequency (or wavelength) independent density turbulence (e.g., \citealt{readhead1971}). 
Figure~\ref{fig_8} (right) shows the plot of $m\cdot\nu$, for the source B0316+162 as a 
function of heliocentric distance. The linear fit in the weak-scattering region extends 
from large to small solar offsets, i.e., from P- to S-bands, and the collective 
radial dependence remains the same as observed at each individual band.

\begin{figure}[t]
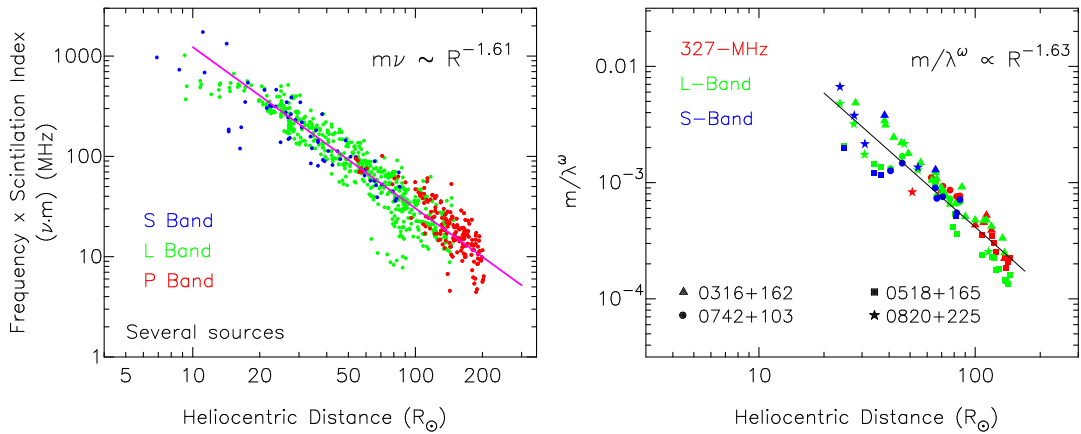

   \centerline{\hspace*{0.00\textwidth}
               \includegraphics[width=5.6cm,angle=-90.0]{figure_9A_new.ps}
               \hspace*{0.001\textwidth}
               \includegraphics[width=5.6cm,angle=-90.0]{figure_9B_new.ps} 
              }
   \vspace{0.02\textwidth}
\caption{
(Left) Similar to Figure \ref{fig_8}, the `$m\cdot\nu$ - $R$' plot of all sources 
observed with Arecibo's P-band (red dots), L-band (green dots), and S-band (blue dots) 
systems. The least-square fit to the observed data is shown as a straight line. 
(Right) The `$m/\lambda^\omega \sim R$' plot for the four sources shown in Figure 
\ref{fig_6}. Each source's symbol is indicated, and the frequency-band color code is
the same as in the left plot. 
}
\label{fig_9}
\end{figure}

Similar to the B0316+162 plot, Figure \ref{fig_9} (left) presents the `$m\cdot\nu$ - $R$' 
plot for all sources observed with the three Arecibo bands. The best linear fit for the
measurements at three bands demonstrates the consistency in the radial dependence. 
It is also in agreement with the result of multiple-source observations presented 
in Figure 2 of \citet{readhead1971}, $m\cdot\nu \propto R^{-1.59}$.
Figure \ref{fig_9} (right) presents a plot of `$m/\lambda^\omega \sim R$' for the 
four sources shown in Figure \ref{fig_6}. The radial index exhibits consistent behavior for
these sources.  However, the data points associated with B0518+165 lie below the best-fit 
line, likely due to its larger angular size compared to the other sources considered.

\begin{figure}[t]
   \centerline{
   \includegraphics[width=7.7cm,angle=-90.0]{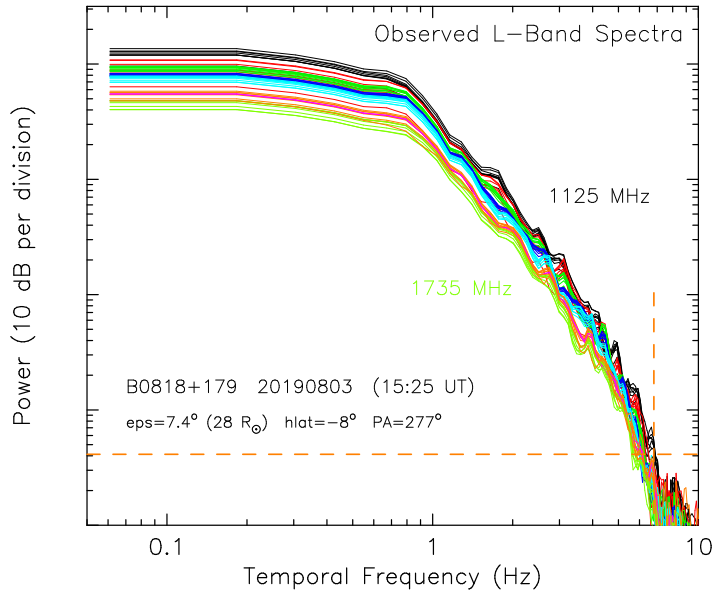} }
   \vspace{0.02\textwidth}
\caption{
Temporal power spectra of 56 channels at L-band, between 1125 and 1735 MHz, for the
source B0818+179 on 3 August 2019. Each spectrum is separated by 10 MHz. 
Spectra within an 80-MHz band, (eight channels corresponding to one
Mock box), are represented by a single color.
The horizontal 
dashed line indicates the spectral off level where the scintillation power drops to 
the background noise level, while the vertical dashed line indicates the frequency at 
which this occurs.  The off-level spectrum has been subtracted from the power spectrum.
} 
\label{fig_10}
\end{figure}

\section{Frequency Dependence of the Temporal Power Spectrum}

\subsection{L-band Observations}

Following the systematic decreasing trend observed in the `$m$ -- $\lambda$' relationship
(see Section 4.1 and Figures \ref{fig_3} to \ref{fig_5}), a comprehensive 
analysis has been performed to investigate the evolution of the temporal spectrum at
the three bands of the Arecibo system. Figure \ref{fig_10} displays the shapes 
and power levels of the temporal power spectra between 1125 and 1735 MHz for the source 
B0818+179. The observations were taken on 3 August 2019, at a solar 
elongation, $\varepsilon$~=~7.4$^\circ$ (R~$\approx$~28~$R_\odot$), when the Sun--P-point position
angle was 277$^\circ$. The position angle, PA, is measured counterclockwise from the north 
pole.   Each spectrum corresponds to a 10-MHz channel, and a total of 
56 spectra are plotted. The maximum of each spectrum has been normalized to that of the
spectrum at the lowest 
frequency, 1125 MHz, and the power level decreases from 1125 to 1735 MHz. 
Spectra within an 80-MHz band, (eight channels corresponding to one Mock box), are represented 
by a single color. A vertical dashed line marks the frequency at which the scintillation 
equals the system noise. Above this `cutoff' frequency, the spectrum is nearly flat, resembling 
white noise. The horizontal line indicates the white-noise level, which has been subtracted 
from the power spectrum.

\begin{figure}[t]
   \centerline{
   \includegraphics[width=7.7cm,angle=-90.0]{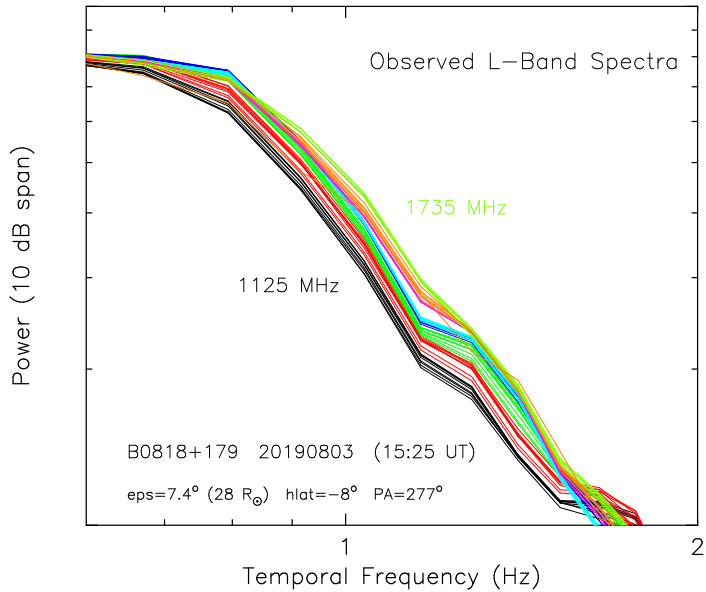} }
\caption{
Progressive broadening of the temporal spectrum with the observing frequency. 
Each spectrum plotted in Figure \ref{fig_10} has been normalized to a maximum level
and plotted on an enlarged scale to show the increasing width of the spectrum with the 
frequency of observation.
The color scheme is same as Figure \ref{fig_10}.
} 
\label{fig_11}
\end{figure}

In the above spectra, the flat spectral part at temporal frequencies below 0.7 Hz is 
caused by the rising part of the first Fresnel oscillation of the propagation filter 
combined with the power-law form of the spatial spectrum of density turbulence, given by 
$\sin^2\left({\kappa^2}/{\kappa^{2}_{diff}} \right)$ $\times$ $\left (\kappa_{x}^{2}
+ \frac{\kappa_{y}^{2}}{{\rm AR}^2} \right )^{-\alpha /2}$
(refer to Equation~(2) in Section 3.3). 
The sharp `knee-like' feature following the flat portion, where most of the
turbulent power is contained, is known as the `Fresnel knee'. It is caused by the first 
minimum of the Fresnel oscillation. Integration along the line of sight tends to 
smooth out the higher-order Fresnel oscillations. It also alters the systematic slope at 
frequencies above the knee, i.e., the inertial part of the spectrum, to a power-law index 
of $\alpha$--1 (e.g., \citealt{mano1994}).

The anisotropic turbulence structure (axial ratio, AR~$>$~1), if present, can
reduce the amplitude of Fresnel oscillations, causing the `knee' region to become
rounded. The visibility function of the source, $\exp[-(\kappa z \Theta/2.35)^2]$, 
and the dissipation or inner-scale size of the turbulence ($S_i$), 
 $\exp(-\kappa^2/{\kappa_{i}^{2}})$, where $\kappa_i$ = 3/$S_i$, 
effects become noticeable only at high temporal frequencies. For instance, a 
source with an angular width of $\Theta$ = 30~mas attenuates spatial wavenumbers 
$\kappa_{source}~>~0.1$~km$^{-1}$, resulting in reduction of power at temporal 
frequencies $>$~7 Hz for a solar wind velocity of 400 km~s$^{-1}$. At the current 
heliocentric distance, 28 $R_\odot$, the inner-scale cutoff will also be comparable 
to the source-size effect (\citealt{mano1987}; \citealt{colesharmon1991}).

Since the frequency of the knee is linearly proportional to the solar wind velocity, 
expressed as $f_{knee} = V_{SW}/\sqrt{\pi\lambda z}$, the knee position shifts inward or 
outward along the frequency axis with a decrease or increase of the solar wind velocity. 
When the velocity remains constant, the temporal power spectra are expected to broaden 
progressively with observing frequency. This effect is clearly demonstrated in Figure 
\ref{fig_11}, where the spectra are normalized to the maximum power level. This provides 
an enlarged view of the spectra, particularly in the region of the knee, and demonstrates
the spectral broadening.

\begin{figure}[t]
\centerline{
\includegraphics[width=8.4cm,angle=-90.0]{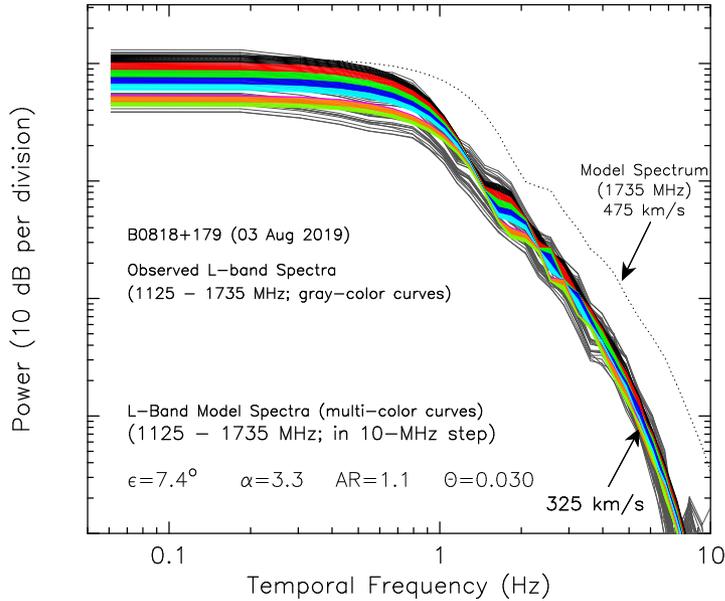} }
\caption{
Set of best-fitted model spectra computed using Equation (2) plotted in multi-color for 
a solar wind velocity of $V_{SW} = 325$~km~s$^{-1}$. In the background, the observed spectra are 
plotted in `gray color'.  The model spectra are plotted with the same color scheme followed 
in Figure \ref{fig_10}. To demonstrate the solar wind velocity scaling of temporal spectrum, 
a sample model spectrum at an observing frequency of 1735 MHz, for $V_{SW} = 475$~km~s$^{-1}$, is 
shown in a dotted line. Model parameters used in the computation are indicated. 
}
\label{fig_12}
\end{figure}

\subsection{L-Band Model Temporal Spectra}

The best-fit model spectra for the L-band spectra displayed in Figure \ref{fig_10} have 
been obtained using Equation (2) and are displayed in Figure \ref{fig_12}. 
For each 10-MHz band, the fitted parameters are: solar wind velocity, $V_{SW}$ = 325 km~s$^{-1}$, 
power-law index, $\alpha$ = 3.3, axial ratio, AR = 1.1, and source size, $\Theta$ = 30~mas.
Model spectra are plotted using the same color scheme as for the observed spectra 
in Figure \ref{fig_10}. 
For comparison, the observed spectra are replotted in gray color in the background. 
Overall, a good agreement is found between the observed and model spectra, with the latter 
closely reproducing key features such as power levels, Fresnel knees, oscillations, and 
temporal-frequency scaling. However, a systematic deviation is observed at the knee 
region. Specifically, compared to the narrower frequency spread of the model knees, the 
observed spectra at low L-band frequencies are shifted towards higher temporal frequencies, 
while high-frequency observed spectra are shifted towards lower temporal frequencies.

\subsubsection{Effective Width of the Temporal Spectrum -- Spectral Moments}

The width of the observed temporal spectrum, especially around the knee, has been carefully 
analyzed and compared to the model spectrum. The first moment, i.e., the area under 
the spectrum, represents the total variance integrated over all temporal frequencies, 
$\sigma^2 = {\int_{0}^{f_c}\,P(f)\, {\rm d}f}$,
where $f_c$ is the cutoff frequency at which the scintillation power roughly drops to 
the noise level of the receive system (refer to Figure \ref{fig_10}).  Higher-order 
moments, normalized by $\sigma^2$, provide information about effective  spectral widths 
at different power levels. The n${^{\rm th}}$-order spectral moment is defined as, 
\begin{equation}
f{_{n^{\rm th}}^{n}}  =  {\frac{1}{\sigma^2}} {\int_{0}^{f_c}\,f^n P(f)\, {\rm d}f}.
\end{equation}
The second and third moments are particularly useful for understanding the spectral
width within about the 2 -- 5 dB down level of the spectrum, where the Fresnel knee is 
prominent, containing the most turbulent power, and can be used to infer the underlying 
physical processes affecting  spectral broadening. 

\begin{figure}[t]
\centerline{
\includegraphics[width=6.5cm,angle=-90]{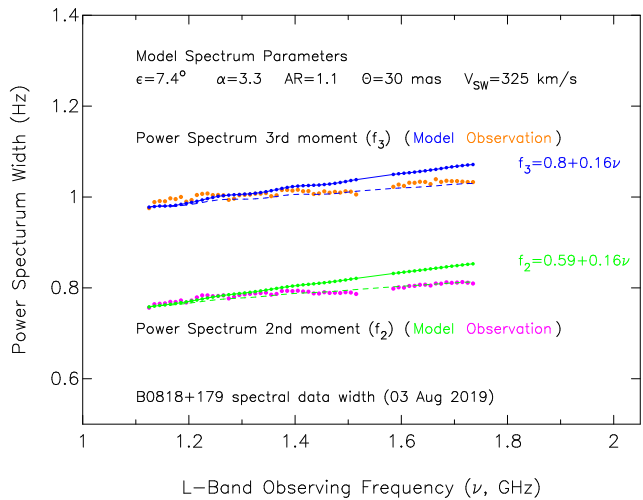} }
\caption{Spectral widths (i.e., second ($f_2$) and third ($f_3$) moments) of the observed spectra, 
for the 56 channels of the L-band system (i.e., the spectra shown in Figure \ref{fig_10}) are  
plotted  against the observing frequency, $\nu$ in GHz. For comparison, widths of model spectra
(shown in Figure \ref{fig_12}) and spectra with progressively increasing axial ratio (see text) 
are plotted. 
}
\label{fig_13}
\end{figure}

In Figure \ref{fig_13}, the second and third moments calculated using Equation (3) 
for the 56-channel model spectra are plotted 
against the observing frequency, $\nu$ (in GHz) as green and blue dotted lines, respectively.
A systematic broadening of the effective 
spectral width with increasing observing frequency is clearly seen. Both the second and 
third moments exhibit linear relationships with the same slope of $\sim$0.16 but differ 
by a constant offset of 0.21 Hz.  The 3-dB down widths of the spectra fall between the 
curves of the second and third moments. The second-moment widths of the observed spectra, 
shown as `pink' dots, align with the model spectra at lower observing frequencies, but
systematically deviate toward lower widths at higher frequencies. Similar deviations 
are also noted in the third-moment widths, represented by `orange' dots.

Since the model spectra were computed with a fixed axial ratio of  AR = 1.1, the 
systematic deviation of the spectral width toward lower temporal frequencies with 
increasing observing frequency suggests that the axial ratio increases with the 
observing frequency. Another set of model spectra was computed, considering a linear 
increase in the axial ratio of about 20\% over the frequency range of the L-band, while 
keeping the other model parameters unchanged. The spectra of the second and third moments, 
obtained from Equation (3), are plotted as green and blue dashed lines in Figure 
\ref{fig_13}. At lower observing frequencies, the spectral widths align with the model 
spectra for the 
fixed axial ratio, and the gradual deviation of the moments with observing frequency 
agrees with the observed spectral moments. The results showing an increase in axial 
ratio with observing frequency (or a decrease with observation wavelength) suggest 
that anisotropy increases as the diffraction-scale size, $\sqrt{\lambda z}$, becomes 
smaller. Conversely, larger turbulent structures tend to be isotropic.


\subsection{Temporal Spectra of Three Bands}

\begin{figure}[t]
\centerline{\includegraphics[width=7.5cm,angle=-90.0]{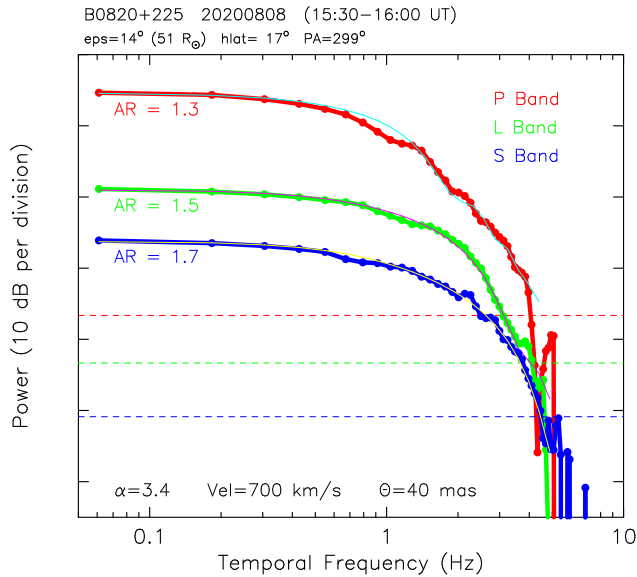} }
\caption{Average spectra observed at P-, L-, and S-bands for B0820+225 on 8 August 2020 
at a heliocentric distance of 51 $R_\odot$ ($\varepsilon$ = 14$^\circ$). The horizontal 
dashed lines are the off-level spectra, corresponding to each band, which have been 
subtracted from the appropriate spectrum. The thin line overplotted on each spectrum is the 
best-fit model spectrum, and for which model parameters are indicated. 
}
\label{fig_14}
\end{figure}

Figure \ref{fig_14} presents the temporal spectra of the source B0820+225 
for the three Arecibo bands, observed between 15:30 -- 16:00~UT 
on 8 August 2020. The source had an elongation of $\varepsilon = 14^\circ$, 
corresponding to a heliocentric distance of 51~$R_\odot$, with a  PA of 
299$^\circ$, and a heliographic latitude of 17$^\circ$. Each 
spectrum, shown as a thick-dotted line, represents the average of the band 
after subtracting its off-level spectrum, indicated by a dashed line.

As exhibited by the `$m$ -- $\lambda$' plot for this source in Figure 
\ref{fig_6}, a clear reduction in scintillation power of $\sim$20 dB 
($P(f) \propto m^2$) between the P and S bands is revealed in the above spectra. 
The spectrum for each band 
has been fitted with the model spectrum obtained using Equation (2) and the 
best-fit model is over-plotted as a thin line on the observed spectrum. 
Model parameters include the power-law index of $\alpha$ = 3.3, a solar-wind 
velocity of $V_{SW}$~=~700 km~s$^{-1}$ and a source angular size of $\Theta$ = 40 mas.
However, the fitted axial ratios for the P-, L-, and S-bands are 1.3, 1.6, and 1.7,
respectively. This progressive increase in axial ratio from low to high 
observing frequency indicates the consistency with the L-band results (Figures 
\ref{fig_12} and \ref{fig_13}), confirming that turbulence scales become more 
anisotropic at smaller scales. 
The fitted solar wind velocity of $\sim$700 km~s$^{-1}$ is consistent with the foot-point 
location of point `P' on a low-emitting large unipolar region and its embedded 
coronal hole, which persisted for several days around the $\pm$30$^\circ$ latitude region 
of the solar equator (refer to the SDO/AIA images in the 211~{\AA} channel about a day before the observation 
date, allowing the solar-wind propagation time to the `P' point; e.g.,
\url{https://sdo.gsfc.nasa.gov/assets/img/browse/2020/08/07/20200807_084811_2048_0211.jpg}).

\begin{figure}[t]
\centerline{\includegraphics[width=7.5cm,angle=-90.0]{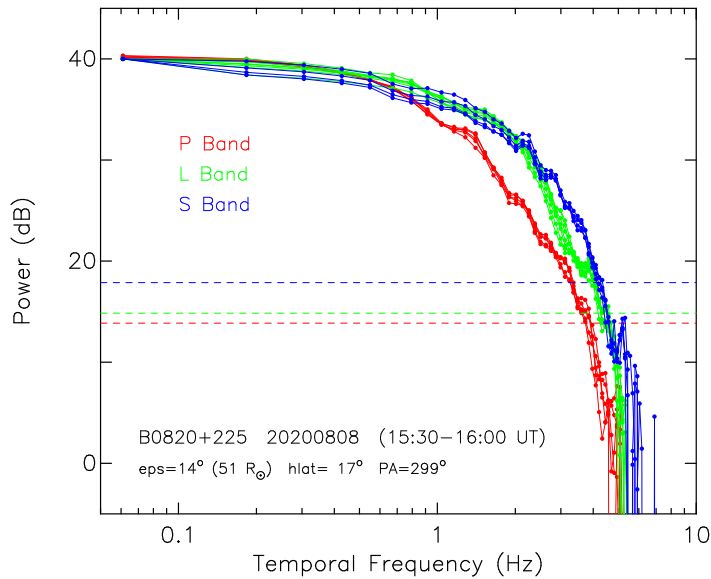} }
\caption{Spectra of the source B0820+225, observed at three bands. 
}
\label{fig_15}
\end{figure}

To examine spectral broadening between bands, the spectra have been normalized to 
a common power level and are displayed in Figure \ref{fig_15}. For the P band, 
individual 10-MHz spectra are shown. For the L and S bands, average spectra 
from 80-MHz bands (each Mock box corresponds to eight channels, each of 10-MHz bandwidth) 
are plotted. The horizontal dashed lines indicate the average off-level spectra
and each of it has been subtracted from the corresponding plotted spectrum. The temporal-frequency broadening 
between the P and L bands is evident over the entire frequency range beyond the 
Fresnel knee. However, due to increased rounding of the knees at the L and S bands, 
only marginal broadening between these is seen at approximately the 15-dB down 
level.

\section{Summary and Conclusions}

Interplanetary scintillation observations obtained from the Arecibo 305-m 
Radio Telescope with the P-, L-, and S-band systems in the frequency range 
of $\sim$300 -- 3100 MHz have been analyzed. These observations covered a
heliocentric distance range of $\sim$5 -- 200 $R_\odot$ in the minimum 
phase at the end of solar cycle 24 and the beginning of cycle 25.

The scintillation dynamic spectrum obtained from L-band observations over a 
frequency range of $\sim$600 MHz permits the tracking of systematic 
reductions in density turbulence within this continuous band. The observed 
level of scintillation at a given frequency is closely associated with the density 
turbulence present in the corresponding micro-scale structures of the solar wind.
The result agrees with the IPS dynamic spectra observed with LOFAR in the frequency 
range of 210 -- 250 MHz, as well as simulated spectra 
(\citealt{fallow2013}; \citealt{coles1984}). 
However, when dynamic spectra were observed in the strong scattering region,
the aforementioned systematic trend in scintillation was not clear 
(\citealt{coleslee1980}; \citealt{hewish1989}; \citealt{fallow2013}).

The dependence of the scintillation index, $m$, on the wavelength of 
observation, $\lambda$, has been quantitatively examined based on the 
estimation of $m$, at frequency intervals of 10 MHz, for many
sources at the L band, along with near-simultaneous observations 
of selected sources covering the P-, L-, and S-band systems. There is a
systematic dependence of $m \sim \lambda^\omega$.
For a given source, the index, $\omega$, remains relatively constant for 
observations made at different heliocentric distances within the 
weak-scattering limit. The values of $\omega$ fall between $\sim$1 and 1.8
and this range is based on observations of several sources. For a given
source, the day-to-day variation of $\omega$ is primarily due to the changes
of density turbulence along different lines of sight (refer to Figures
\ref{fig_3} to \ref{fig_6}). However, the average value of $\omega$ of a source, 
obtained from observations taken on different days, typically represents the 
angular size associated with the source.


The consistency of the `$m$ -- $\lambda$' dependence with the distance 
from the Sun allows a study of the solar wind density turbulence in
the near-Sun region, extending as close as $\sim$10 $R_\odot$,
particularly when using S-band observations. Since the rms of density 
fluctuations, $\delta n_e$ is related to 
{$m^2 (\theta_o,\lambda) \sim \lambda^{2w} \int_z \delta n_{e}^{2}(R,\lambda,z,\theta_o)\;{\rm d}z$.}
Thus, the level of rms density fluctuations giving rise to 
scintillation is controlled by the density scale size relative
to the Fresnel scale present in the solar wind. 
{Since in the weak-scattering region, the $\delta n_e$ and $m$ are linearly 
related, the high-frequency observations allow to study the correlation between
them in the near-Sun region}.

The reduction of scintillation with the wavelength of observation has been
reported in earlier multi-frequency IPS studies, which also include the 
decrease of correlation observed with dual-frequency measurements
(e.g., \citealt{gapper1981}; \citealt{scott1983}; \citealt{coles1984}; 
\citealt{bourgois1985}; \citealt{andy2006}; \citealt{fallows2006}; 
\citealt{liu2010}; \citealt{fallow2013}; \citealt{morgan2017}).

The present study provides the radial evolution of $m$ with 
heliocentric distance at three separate bands. The shifting of the
weak-to-strong scattering transition region closer to the Sun from
low to high frequency is clearly illustrated in Figures 
\ref{fig_7} to \ref{fig_9}. 
Specifically, high-frequency observations effectively allow us to probe 
the solar wind closer to the Sun, where the linear relationship 
between $m$ and $\delta{n_e}$ is maintained. 
The invariance of the
`$m$ -- $\lambda$' relationship with the heliocentric distance has been 
effectively demonstrated by the $m$ measurements at each individual 
band (also $m$ scaled with frequency of observation, $m\cdot \nu$ 
or $m/\lambda$),
Figures \ref{fig_8} and \ref{fig_9}. Such 
multi-frequency IPS measurements are relatively rare in previous 
studies (e.g., \citealt{cohen1969}; \citealt{readhead1978}). 
During the current solar cycle minimum, the observed radial variation 
of $\delta  n_{e}^{2}(R) \sim R^{-4.2}$ is consistent with earlier 
results (\citealt{mano1993}; \citealt{asai1998}; \citealt{mano2012}).

In addition to the `$m$ -- $\lambda$' relationship, the temporal power
spectrum analysis, for 10-MHz intervals of the L-band IPS, indeed
for all three P-, L-, and S-bands, has shown a systematic 
decrease of the power level and frequency broadening of the spectrum with 
increasing observing frequency. The best-fit model to the observed 
spectrum allowed us to infer the typical solar wind velocity and the spatial 
spectrum of density turbulence, i.e., power-law index, $\alpha$. However,
the model-fitting procedure revealed a systematic deviation of the Fresnel 
knee of the spectrum, toward the low frequency, with the observing wavelength. 
Careful
examination indicates that the spectrum involved with the large-scale size
has been associated with more isotropic turbulence scale than the 
small-scale spectrum. The axial ratio increases as the scale size of 
turbulence responsible for the scintillation decreases. 

The above result is consistent with the increase of field-aligned 
anisotropy as the Sun is approached, at heliocentric distance $<$20 
$R_\odot$, where the scale size decreases with solar offset 
(e.g., \citealt{armstrong1990}; \citealt{grall1997}).
Additionally, a recent investigation of the solar wind at 1 AU, based on
the Advanced Composition Explorer (ACE) mission {\it in-situ} data, showed 
that the density correlation 
scale in the direction quasi-parallel to the mean magnetic field is 
slightly larger than that in the quasi-perpendicular direction 
(\citealt{wang2024}). Moreover, some of the simulations and theoretical
studies have confirmed the increased anisotropy with decreased scale size  
(e.g., \citealt{shebalin1983}; \citealt{beresnyak2005}; 
\citealt{oughton2005}).
The multi-frequency IPS observations presented in this study highlight 
the significance of probing the characteristics of solar wind density 
turbulence at various spatial scales and distances from the Sun.

During these multi-frequency IPS observing sessions, some weak CME events 
were also detected. These observations are valuable for understanding
CME propagation in the inner heliosphere. Additionally, the high sensitivity 
of the Arecibo telescope provided spectra extending to high temporal 
frequencies, enabling estimation of the inner scale of turbulence 
for sources with known VLBI source-size structures. The results of these 
investigations will be presented elsewhere.


\begin{acks}

We thank the observational, computational, and engineering support provided 
by Phil Perillat, Arun Venkataraman and other staff members of the Arecibo 
Observatory. 
The Arecibo Observatory was operated by the University of Central Florida
under a cooperative agreement with the National Science Foundation, and in 
alliance with Universidad Ana G. M{\'e}ndez and Yang Enterprises, Inc.  
We also thank the observing team of the Radio Astronomy Centre 
(RAC) for the 327-MHz IPS observations. The RAC is run by the National Centre 
for Radio Astrophysics of the Tata Institute of Fundamental Research, India. 
PKM wishes to thank Tapasi Ghosh for numerous useful discussions and suggestions 
during the preparation of the manuscript. 

\vspace{0.04\textwidth}

{\noindent {\bf Funding} 
The Arecibo Observatory was operated by the University of Central Florida
under a cooperative agreement with the National Science Foundation (grant 
number: AST-1822073), and in alliance with Universidad Ana G. M{\'e}ndez 
and Yang Enterprises, Inc. PKM acknowledges support from the University of 
Central Florida. He also acknowledges support from NASA GSFC through the
Cooperative Agreement to the Catholic University of America in support of 
the Partnership for Heliophysics and Space Environment Research (PhaSER) 
under the grant 80NSSC21M0180.  CJS did not receive funding for this work. 
}

\vspace{0.04\textwidth}

{\noindent {\bf Data Availability}
The observed IPS datasets analyzed in the current study are available at the Arecibo
Observatory data archive maintained at the Texas Advanced Computing Center 
(www.tacc.utexas.edu/about/help/). The temporal power spectra generated during the 
current study are available from the corresponding author on request. 
}
\vspace{0.04\textwidth}

{\noindent {\large\bf Declarations}
\vspace{0.003\textwidth}

{\noindent \bf Competing interests} The authors declare no competing interests.

}

\end{acks}


\appendix
\renewcommand\thefigure{\thesection.\arabic{figure}}    
\setcounter{figure}{0}
\section{Supplementary Figures}

This appendix contains some additional figures to support the results described 
in Section 4. 

\begin{figure}[h]
\centerline{\includegraphics[width=7.0cm,angle=-90.0]{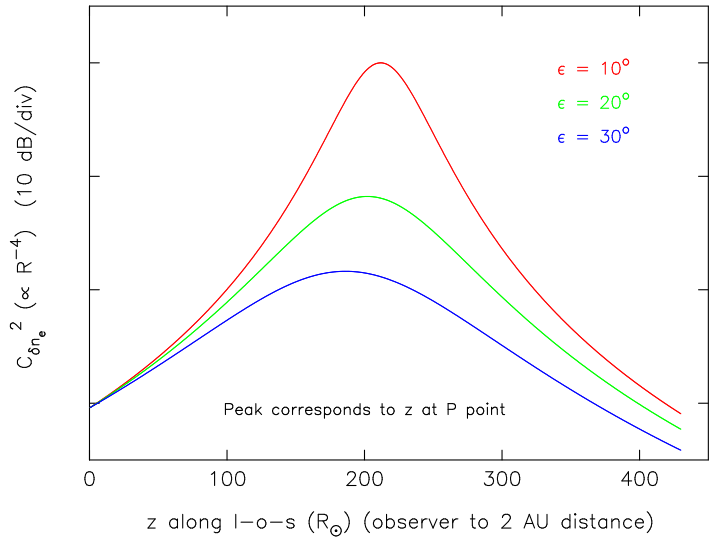} }
\caption{
Scattering level plotted as a function of distance, {\it z} (along the line of sight), from the observer 
to 2 AU, on a `log-linear' scale. The plots are shown for three elongations as indicated. 
For these plots, the $\beta$ value of $-$4 has been used in the power-law dependence 
of $C^{2}_{\delta n_e}(R) \propto R^\beta$. For each elongation, the peak in the 
scattering level corresponds to the location of the solar-wind layer at the `P' point.
}
\label{fig_a1}
\end{figure}

\begin{figure}[h]
\centerline{\includegraphics[width=7.0cm,angle=-90.0]{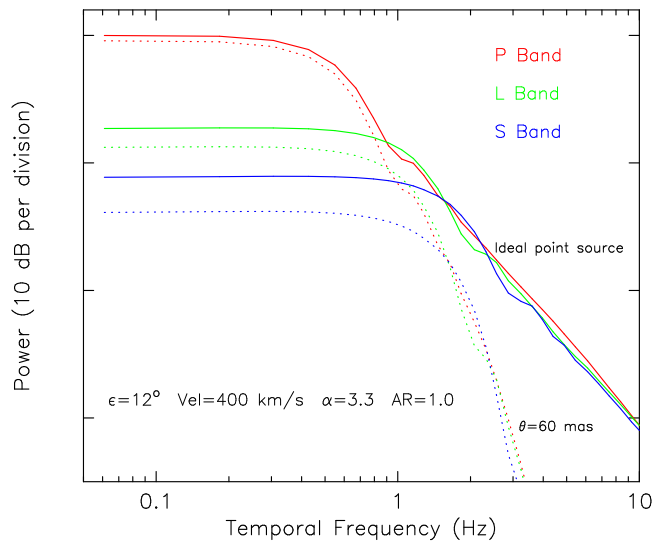} }
\caption{
Model spectra at the middle of the P-, L-, and S-bands, computed using Equation (2).
Solid lines and dotted lines represent an ideal point source and a source with a size 
of 60~mas, respectively. The model parameters used in the computation are indicated.
In this log-log plot, the difference in power level between the point source spectrum 
and the finite source size spectrum progressively increases with decreasing wavelength 
(or increasing frequency). The derived $\omega$ values are 1.08 for the point source 
and 1.4 for the finite source.
}
\label{fig_a2}
\end{figure}

\newpage

\bibliographystyle{spr-mp-sola}
\bibliography{manuscript}

\end{document}